\documentclass[prd,aps,amsmath,showpacs,nofootinbib]{revtex4}

\usepackage{times}
\usepackage{hyperref}

\usepackage{graphicx}
\usepackage{dcolumn}
\usepackage{bm}
\usepackage{color} 

\begin{document}

\title{
Four-body Dalitz plot contribution to the radiative corrections in $K_{l3}^0$ decays and its role in the determination of $|V_{us}|$
}

\author{
M.\ Neri
}
\affiliation{
Departamento de F{\'\i}sica, Escuela Superior de F\'{\i}sica y Matem\'aticas del Instituto Polit\'ecnico Nacional, Apartado Postal 75-702, Ciudad de M\'exico 07738, Mexico
}

\author{
A.\ Mart{\'\i}nez
}
\affiliation{
Departamento de F{\'\i}sica, Escuela Superior de F\'{\i}sica y Matem\'aticas del Instituto Polit\'ecnico Nacional, Apartado Postal 75-702, Ciudad de M\'exico 07738, Mexico
}

\author{
C.\ Ju\'arez-Le\'on
}
\affiliation{
Departamento de F{\'\i}sica, Escuela Superior de F\'{\i}sica y Matem\'aticas del Instituto Polit\'ecnico Nacional, Apartado Postal 75-702, Ciudad de M\'exico 07738, Mexico
}

\author{
J.\ J.\ Torres
}
\affiliation{
Departamento de Posgrado, Escuela Superior de C\'omputo del Instituto Polit\'ecnico Nacional, Apartado Postal 75-702, Ciudad de M\'exico 07738, Mexico
}

\author{
Rub\'en Flores-Mendieta
}
\affiliation{
Instituto de F{\'\i}sica, Universidad Aut\'onoma de San Luis Potos{\'\i}, \'Alvaro Obreg\'on 64, Zona Centro, San Luis Potos{\'\i}, San Luis Potos{\'\i} 78000, Mexico
}

\date{\today}

\begin{abstract}
The four-body contribution of the model-independent radiative corrections to the Dalitz plot of the semileptonic decays of neutral kaons are computed to order $(\alpha/\pi)(q/M_1)$, where $q$ is the momentum transfer and $M_1$ is the kaon mass. The final result is presented in two forms. The first one is given in terms of the triple integration of the bremsstrahlung photon ready to be performed numerically; the second one is a fully analytical expression. This paper is organized to make it accessible and reliable in the analysis of the Dalitz plot of precision experiments involving kaons and is not compromised to fixing the form factors at predetermined values. As a byproduct, gathering together three- and four-body contributions of radiative corrections yields, through a least-squares fit to the measured kaon decay rates, the value $f_+^{K^0\pi^-}|V_{us}| = 0.2168(3)$.
\end{abstract}


\maketitle

\section{Introduction}

Kaon semileptonic ($K_{l3}$) decays play a leading role in the determination of the Cabibbo-Kobayashi-Maskawa (CKM) matrix element $|V_{us}|$. Unlike baryon semileptonic decays where both vector and axial-vector currents participate with the introduction of six form factors, $K_{l3}$ decays are well described with only two vector form factors. An extra complexity is added by SU(3) symmetry-breaking effects unavoidably present in these form factors. While there are large uncertainties due to first-order symmetry-breaking corrections in the axial-vector form factors, the vector ones are protected by the Ademollo-Gatto theorem \cite{ag} against SU(3)-breaking corrections to lowest order in $(m_s-\hat{m})$. As a result of this fortunate situation, the accuracy on $|V_{us}|$ is approaching the 1\% level.

An important source of systematic errors of theoretical nature also to be considered in a precise determination of $|V_{us}|$ is the inclusion of radiative corrections (RC). The RC to the Dalitz plot (DP) of $K_{l3}$ decays have been dealt with in previous works \cite{jl2011,jl2012,jl2015} to order $(\alpha/\pi)(q/M_1)$, where $q$ is the momentum transfer and $M_1$ is the kaon mass. Specifically, for charged kaons the so-called three- and four-body regions of this DP were considered in Refs.~\cite{jl2011} and \cite{jl2012}, respectively, whereas for neutral kaons only the three-body region was considered in Ref.~\cite{jl2015}. Hereafter, these regions will be loosely referred to as TBR and FBR, respectively. The distinction between these two regions is discussed in detail in Ref.~\cite{jl2011} so unnecessary repetitions will be avoided here; suffice it to say that the FBR is present when real photons cannot be discriminated in an experimental analysis. Thus, in the calculation of bremsstrahlung RC a clear distinction between these two regions should be kept.

The aims of the paper are twofold. The first one is to extend the analysis of Ref. \cite{jl2015} in order to calculate the FBR contribution of the RC to the DP of $K_{l3}^0$ decays, on the same footing as Ref.~\cite{jl2012}; this completes the calculational program about RC in $K_{l3}$ decays initiated in Ref.~\cite{jl2011}. The second and most important one is finally the implementation of the full RC of order $(\alpha/\pi)(q/M_1)$ towards the determination of $|V_{us}|$. A word of caution is in order. It is assumed here that the form factors can be extracted from either experiment or lattice QCD simulations or by another approach, and thus they are not considered in the analysis; their model-independent parts contain information only on QED. Under this premise, the determination of $|V_{us}|$ can be unambiguously achieved.

The organization of the paper is as follows. In Sec.~\ref{sec:fbr} the notation and conventions are first settled down; next the triple integrals over the bremsstrahlung photon are properly identified and are left ready to be performed numerically. Gathering together partial results yields a preliminary expression for the FBR contribution to the DP. In Sec.~\ref{sec:ai} those integrals are computed analytically using general properties of the integrands under rotations. This is indeed a remarkable achievement because it allows one to construct a fully analytical expression for the DP of $K_{l3}^0$ decays including RC of order $(\alpha/\pi)(q/M_1)$. In Sec.~\ref{sec:vus} the RC are incorporated into the total decay rate of $K_{l3}$ decays by collecting all partial results of previous works \cite{jl2011,jl2012,jl2015}. At this stage, a numerical comparison between the theoretical expressions so obtained and the available experimental data \cite{particle} is possible in order to extract information about the only free parameter left, namely, the product $f_+^{K^0\pi^-}(0)|V_{us}|$. Thus, a proper determination of $f_+(0)$ would result in a precise value of $|V_{us}|$ from $K_{l3}$ decays. In Sec.~\ref{sec:summ} a summary and some closing remarks are provided.

\section{\label{sec:fbr}The four-body contribution of RC in $K_{l3}^0$ decays}

The bremsstrahlung RC is a four-body decay whose DP covers entirely the three-body decay
\begin{equation}
K^0(p_1) \to \pi^-(p_2) + l^+(l) + \nu_l(p_\nu^0). \label{eq:tbd}
\end{equation}
The notations and conventions can be found in Ref.~\cite{jl2011}. Put succinctly, the four-momenta of $K^0$, $\pi^-$, $l^+$, and $\nu_l$ are $p_1=(E_1,\mathbf{p_1})$, $p_2=(E_2,\mathbf{p_2})$, $l=(E,\mathbf{l})$, and $p_\nu^0=(E_\nu^0,\mathbf{p_\nu^\mathrm{0}})$. The nonzero masses of the first three particles are $M_1$, $M_2$, and $m$, respectively.

For definiteness, the four-body decay to be considered here is denoted by
\begin{equation}
K^0(p_1) \to \pi^-(p_2) + l^+(l) + \nu_l (p_\nu) + \gamma (k), \label{eq:fbd}
\end{equation}
where $\gamma$ represents a real photon with four-momentum $k=(\omega,\mathbf{k})$. Simple relations can be obtained for the energy and three-momentum of the neutrino for processes (\ref{eq:tbd}) and (\ref{eq:fbd}), namely, $E_\nu = E_\nu^0-\omega$ and $\mathbf{p_\nu} = \mathbf{p_\nu^\mathrm{0}}-\mathbf{k}$. Furthermore, if the calculation is performed in the center-of-mass frame of the kaon, then $M_1=E+E_2+E_{\nu}+\omega$ and $\mathbf{0}=\mathbf{p}_2+\mathbf{l}+\mathbf{p}_\nu + \mathbf{k}$.

The calculation of bremsstrahlung in the FBR is rather straightforward because the amplitude $\mathsf{M}_B$ keeps the same structure as the one given for the TBR in Eq.~(27) of Ref.~\cite{jl2015}, except for the fact that it is now infrared convergent. Only a couple of minor modifications are required. The first one has to deal with replacing the upper limit of the integrals over the variable $y=\hat{\mathbf{p}}_2 \cdot \hat{\mathbf{l}}$ of Eqs.~(64), (67), and (68) of this reference---this limit becomes one---and the second one has to deal with replacing the previously infrared-divergent integral $I_0$ with the convergent one defined as\footnote{Hereafter, the subscript $Fn$ will designate quantities defined in the FBR of the neutral process.}
\begin{equation}
I_{0Fn} = \frac{p_2l}{4\pi} \int_{-1}^1 dx \int_{-1}^1 dy \int_0^{2\pi} d\phi_k \frac{\omega}{D} \left[-\frac{l^2}{(l\cdot k)^2} + \frac{2p_2 \cdot l}{l\cdot k p_2\cdot k} - \frac{p_2^2}{(p_2 \cdot k)^2} \right], \label{eq:i0Fn}
\end{equation}
where $D=E_\nu^0 + (\mathbf{l}+\mathbf{p}_2)\cdot \hat{\mathbf{k}}$.

With these simple changes, the differential decay rate in the FBR becomes
\begin{equation}
d\Gamma_B^\mathrm {FBR}(K_{l3}^0) = \frac{\alpha}{\pi} \frac{G_F^2}{32\pi^3} |V_{us}|^2 M_1^3 dEdE_2 \left( A_0 I_{0Fn} + A_{BFn}^\prime \right), \label{eq:dgbfbr}
\end{equation}
where $G_F$ is the Fermi constant.

The function $A_0$ is already defined in Eq.~(7) of Ref.~\cite{jl2015} whereas the function $A_{BFn}^\prime$ is given by
\begin{equation}
A_{BFn}^\prime = A_{1Fn}^{(B)} |f_+(q^2)|^{2} + A_{2Fn}^{(B)} \mathrm{Re} [f_+(q^2)f_-^*(q^2)] + A_{3Fn}^{(B)} |f_-(q^2)|^2, \label{eq:abFn}
\end{equation}
where
\begin{equation}
A_{1Fn}^{(B)}=\Lambda_{1Fn}+\Lambda_{4Fn}+\Lambda_{7Fn}, \label{eq:a1fn}
\end{equation}
\begin{equation}
A_{2Fn}^{(B)}=\Lambda_{2Fn}+\Lambda_{5Fn}+\Lambda_{8Fn}, \label{eq:a2fn}
\end{equation}
\begin{equation}
A_{3Fn}^{(B)}=\Lambda_{3Fn}+\Lambda_{6Fn}+\Lambda_{9Fn}. \label{eq:a3fn}
\end{equation}

Similarly, the functions $\Lambda_{iFn}$ read
\begin{equation}
\Lambda_{1Fn,2Fn,3Fn} = \frac{p_2l}{4\pi} \frac{8}{M_1^2} \int_{-1}^1 dx \int_{-1}^1 dy  \int_0^{2\pi} d\phi_k \frac{\omega}{D} [b_{11},b_{12},b_{13}], \label{eq:lb1}
\end{equation}
\begin{equation}
\Lambda_{4Fn,5Fn,6Fn} = \frac{p_2l}{4\pi} \frac{8}{M_1^2} \int_{-1}^1 dx \int_{-1}^1 dy \int_0^{2\pi} d\phi_k \frac{\omega}{D} [b_{21},b_{22},b_{23}], \label{eq:lb2}
\end{equation}
and
\begin{equation}
\Lambda_{7Fn,8Fn,9Fn} = \frac{p_2l}{4\pi} \frac{8}{M_1^2} \int_{-1}^1 dx \int_{-1}^1 dy \int_0^{2\pi} d\phi_k \frac{\omega}{D} [b_{31},b_{32},b_{33}], \label{eq:lb3}
\end{equation}
where the $b_{jk}$ factors can be found in Ref.~\cite{jl2015}.

At this point, the nine integrals $\Lambda_{iFn}$, Eqs.~(\ref{eq:lb1})--(\ref{eq:lb3}), and $I_{0Fn}$, Eq.~(\ref{eq:i0Fn}), can be performed numerically to make up Eq.~(\ref{eq:dgbfbr}), which constitutes the first final result for the FBR of the DP of $K_{l3}^0$ decays. However, it is even possible to perform analytically such integrals. In the next section, fully analytical results for this region of the DP will be provided.

\section{\label{sec:ai}Analytical integrations}

The triple integrals pending in the functions $I_{0Fn}$ and $\Lambda_{iFn}$ can be performed analytically by exploiting the transformation properties of the integrands under rotations. In other words, a proper choice of orientation of the coordinate axes notably simplifies the task. A lengthy but standard calculation yields for $I_{0Fn}$
\begin{equation}
I_{0Fn} = I_{0f} (w_+) - I_{0f} (w_-) - 2 \ln \frac{y_0+1}{y_0-1},
\end{equation}
where
\begin{eqnarray}
I_{0f}(w) & = & (\ln w)^2 + \frac{2\alpha_1}{\sqrt{\alpha_1^2-1}}\left\{ \ln w\left[\ln\left(1-\frac{w}{\alpha_1+\sqrt{\alpha_1^2-1}}\right) - \ln\left(1+\frac{w}{-\alpha_1+\sqrt{\alpha_1^2-1}}\right)\right]\right. \nonumber \\
&  & \mbox{} + \left. L\left(\frac{w}{\alpha_1-\sqrt{\alpha_1^2-1}}\right) - L\left(\frac{w}{\alpha_1+\sqrt{\alpha_1^2-1}}\right)\right\}.
\end{eqnarray}
Here $L(x)$ is the Spence function and
\begin{equation}
w_{\pm} = \frac{a+\eta_\pm + \sqrt{(a+\eta_\pm)^2 - 4m^2M_2^2}}{2mM_2},
\end{equation}
\begin{equation}
\eta_\pm = 2p_2l(y_0 \pm 1),
\end{equation}
\begin{equation}
y_0 = \frac{{E_\nu^0}^2-p_2^2-l^2}{2p_2l},
\end{equation}
\begin{equation}
a= 2(EE_2 - p_2ly_0),
\end{equation}
and
\begin{equation}
\alpha_1=\frac{a}{2mM_{2}}.
\end{equation}

Analogously, the fully analytical versions of the functions $\Lambda_{iFn}$ read
\begin{eqnarray}
\frac{M_1^2}{4p_2l} \Lambda_{1Fn} & = & \left[\frac{2E}{M_1}-\frac{m^2}{M_1^2}\right] \left[ M_1(1-\beta^2)\theta_{2F} +\frac{M_1M_2^2}{E_2^2}\theta_{2F}^\prime - 2E_2\theta_{3F} - 2E\theta_{3F}^\prime + \frac{2\zeta_{11F}}{E} + \frac{2\zeta_{11F}^\prime }{E_2} - 4J_{1Fn} \right] \nonumber \\
&  & \mbox{} - \left[\frac{4M_1^2}{m^2}-1\right] \frac{M_1^2}{4p_2l} \Lambda_{3Fn},
\end{eqnarray}
\begin{eqnarray}
\frac{M_1^4}{4p_2lm^2} \Lambda_{2Fn} & = & -\eta_{0F} + M_1(1-\beta^2) \theta_{2F} + \frac{M_1M_2^2}{2E_2^2} \theta_{2F}^\prime - \left[\frac{E}{2}(1-\beta^2) + 2E_2 \right] \theta_{3F} - \left[E-\frac{M_2^2}{2E_2}\right] \theta_{3F}^\prime + 2\frac{\zeta_{11F}}{E} \nonumber \\
&  & \mbox{} + \frac{\zeta_{11F}^\prime}{E_2} - 2J_{1Fn},
\end{eqnarray}
\begin{equation}
\frac{M_1^4}{p_2lm^2} \Lambda_{3Fn} = 2\eta_{0F} + E(1-\beta^2) \theta_{3F} + \frac{M_2^2}{E_2^2} (M_1 \theta_{2F}^\prime - E_2 \theta_{3F}^\prime) - 2E \theta_{3F}^\prime + \frac{2}{E_2} \zeta_{11F}^\prime - 4J_{1Fn}, 
\end{equation}
\begin{eqnarray}
\frac{M_1^2}{4p_2l} \Lambda_{4Fn} & = &  \left[-2E_\nu^0 - \left[1-\frac{m^2}{4M_1^2} \right] \beta p_2y_0 - 3E + \beta l\right] \theta_{3F} + (2E_\nu^0+3E) \theta_{4F} + 3l\theta_{5F} + \frac{E_\nu^0}{E} \theta_{7F} - \frac{1}{2E} \theta_{9F}  \nonumber \\
&  & \mbox{} + \left[1-\frac{m^2}{4M_1^2} \right] \frac{1}{E} \zeta_{11F},
\end{eqnarray}
\begin{equation}
\frac{M_1^2}{4p_2l} \Lambda_{5Fn} = - \frac{m^2}{2M_{1}^{2}} \left[\beta p_2y_0 \theta_{3F} - \frac{1}{E} \zeta_{11F} \right],
\end{equation}
\begin{equation}
\Lambda_{6Fn} = - \frac12 \Lambda_{5Fn},
\end{equation}
\begin{eqnarray}
\frac{M_1^2}{8p_2l} \Lambda_{7Fn} & = & \frac{M_1^2}{4p_2l} \left[\frac{M_1^2}{m^2}-\frac14\right] \Lambda_{8F} - (1-\beta^2)\left[E_{\nu}^0\theta_{2F}-E(\theta_{3F}-\theta_{2F})-\frac{\theta_{6F}}{2}\right] +E_{\nu}^0(\theta_{3F}-\theta_{3F}^\prime) \nonumber \\
&  & \mbox{} - \left[E(\theta_{4F}-\theta_{3F})-E_2(\theta_{4F}^\prime-\theta_{3F}^\prime)+\frac{\theta_{7F}}{2}-\frac{\theta_{7F}^\prime}{2}\right] + \frac{E_{\nu}^0}{M_1}\left[E_2\theta_{3F}+E\theta_{3F}^\prime-\frac{\zeta_{11F}}{E}-\frac{\zeta_{11F}^\prime}{E_2}+2J_{1Fn}\right] \nonumber \\
&  & \mbox{} - \frac{E_2E}{M_1} \left[\theta_{4F}-\theta_{3F}+\theta_{4F}^\prime-\theta_{3F}^\prime+\frac{\theta_{7F}}{2E}+\frac{\theta_{7F}^\prime}{2E_2}\right] - \frac{a}{2M_1^2}\left[\frac{p_2ly_0\theta_{3F}-\zeta_{11F}}{E}+\frac{p_2ly_0\theta_{3F}^\prime-\zeta_{11F}^\prime}{E_2}\right] \nonumber \\
&  & \mbox{} +\frac{\zeta_{21F}}{2M_1E}+\frac{\zeta_{21F}^\prime}{2M_1E_2}-\frac{I_{13F}}{4M_1^2E}-\frac{I_{13F}^\prime}{4M_1^2E_2} + \frac{a}{M_1^2} J_{1Fn} - \frac{EE_2}{M_1^2}J_{2Fn},
\end{eqnarray}
\begin{equation}
\frac{M_1^4}{2p_2lm^2} \Lambda_{8Fn} = 2\eta_{0F} + [E(1-\beta^2)+E_2-M_1] \theta_{3F} + \left[M_1-E-\frac{M_2^2}{E_2}\right] \theta_{3F}^\prime + \frac{\zeta_{11F}^\prime}{E_2} - \frac{\zeta_{11F}}{E} - 2J_{1Fn}, 
\end{equation}
\begin{equation}
\Lambda_{9Fn} = - \frac12 \Lambda_{8Fn},
\end{equation}
with
\begin{equation}
\theta_{0F} = \frac{2}{\beta}\log{\frac{1+\beta}{1-\beta}} - 4,
\end{equation}
\begin{equation}
J_{1Fn} = \frac{2}{\beta\beta_2} (\beta_2 \, \mathrm{arctanh} \, \beta + \beta \, \mathrm{arctanh} \, \beta_2 - \beta \beta_2),
\end{equation}
\begin{equation}
J_{2Fn} = J_{1Fn} + 2\left[\beta\mathrm{arctanh} \,\beta + \beta_2 \mathrm{arctanh} \, \beta_2 + \frac{(1-\beta^2)(1-\beta_2^2)}{\beta\beta_2} \mathrm{arctanh} \, \beta \, \mathrm{arctanh} \, \beta_2 \right],
\end{equation}
and
\begin{eqnarray}
\frac{I_{13F}}{(2p_2l)^2} & = & \frac{E_\nu^0}{p_2^2} \eta_{0F} + \frac{\beta E_\nu^0+l-p_2}{2\beta p_2^2}\theta_{0F} + \frac{3(E+E_\nu^0)^2-l^2-p_2^2-\beta^2(2E+E_\nu^0)E_\nu^0+\beta^2 (1+2 y_0^2)p_2^2}{2\beta^2p_2^2} \theta_{3F} \nonumber \\
&  & \mbox{} - \frac{3 (E+E_\nu^0)^2-l^2-p_2^2-2\beta^2EE_\nu^0}{2 \beta^2 p_2^2}\theta_{4F} - \frac{3(E+E_\nu^0)^2-l^2-p_2^2}{2\beta p_2^2} \theta_{5F} - \frac{3(E+2E_\nu^0)E}{2p_2^2} \theta_{10F} \nonumber \\
&  & \mbox{} - \frac{3(E+E_\nu^0)-\beta p_2}{2\beta p_2} \theta_{12F} + \frac{3(E+E_\nu^0)}{2\beta p_2} \theta_{13F} + \frac{3E}{2p_2}\theta_{19F} - \frac{3El}{2p_2^2} \theta_{20F} - \frac{2y_0}{p_2l} \zeta_{11F}.
\end{eqnarray}

The functions $\theta_{kF}$ ($k=1,\ldots 16$) are given in Ref.~\cite{rfm00} and $\theta_{19F}$, $\theta_{20F}$, $\eta_{0F}$ and $\zeta_{ijF}$ are listed in Ref.~\cite{jj2006}. The primed versions of $\theta_{kF}$ and $\zeta_{ijF}$ are found by making the replacement $p_2\leftrightarrow l$, $E_2 \leftrightarrow E$, and $M_2 \leftrightarrow m$. Also, $\beta=l/E$ and $\beta_2 = p_2/E_2$.

Equation (\ref{eq:dgbfbr}) can be rearranged as 
\begin{equation}
d\Gamma_{B}^{\mathrm{FBR}}(K_{l3}^0) = \frac{\alpha}{\pi} \frac{G_F^2}{32\pi^3}|V_{us}|^2M_1^3 dEdE_2 \left[ A_{1Fn} |f_+(q^2)|^2 + A_{2Fn} \mathrm{Re}[f_+(q^2)f_-^*(q^2)] + A_{3Fn} |f_-(q^2)|^2 \right], \label{eq:dgb2}
\end{equation}
where
\begin{equation}
A_{iFn} = A_{in}^{(0)}I_{0Fn} + A_{iFn}^{(B)}, \qquad \qquad (i=1,2,3).
\end{equation}

The quantities $A_{in}^{(0)}$ correspond to $A_i^{(0)}$ defined in Eqs.~(17)---(19) of Ref.~\cite{jl2011} while $A_{iFn}^{(B)}$ are defined in Eqs.~(\ref{eq:a1fn})---(\ref{eq:a3fn}). Equation (\ref{eq:dgb2}) is a fully analytic expression for FBR of the DP of $K_{l3}^0$ decays and constitutes the second main result of this paper.

The complete DP for the decay $K^0 \to \pi^-l^+\nu_l$, with model-independent RC up to order $(\alpha/\pi)(q/M_1)$, is thus given by
\begin{equation}
d\Gamma (K_{l3}^0) = d\Gamma^{\mathrm{TBR}} + d\Gamma^{\mathrm{FBR}}, \label{eq:dgfinal}
\end{equation}
where $d\Gamma^{\mathrm{TBR}}$ can be found in Eq.~(87) of Ref.~\cite{jl2015} and $d\Gamma^{\mathrm{FBR}}$ is given in Eq.~(\ref{eq:dgb2}).

For completeness, the numerical evaluations of RC for both TBR and FBR are provided in Tables \ref{t:a1ke3} and \ref{t:a1km3} for processes $K^0\to\pi^-e^+\nu_e$ and $K^0\to\pi^-\mu^+\nu_\mu$, respectively.

\begingroup
\begin{table*}
\caption{\label{t:a1ke3}Radiative corrections $(\alpha/\pi)A_1 \times 10$ and $(\alpha/\pi)A_{1F} \times 10$ for $K^0 \to \pi^- + e^+ + \nu_e$ decay. The entries corresponding to the FBR are marked in boldface characters. $E$ and $E_2$ are given in $\textrm{GeV}$.}
\begin{ruledtabular}
\begin{tabular}{crrrrrrrrr}
$E_2$\textbackslash$E$ & $0.0124$ & $0.0373$ & $0.0622$ & $0.0871$ & $0.1120$ & $0.1368$ & $0.1617$ & $0.1866$ & $0.2115$          \\ 
\hline 
\hline
$0.2612$ & $         0.1707$ & $         0.2193$ & $         0.1708$ & $         0.0804$ & $        -0.0262$ & $        -0.1287$ & $-0.2074$ & $-0.2381$ & $-0.1789$ \\
$0.2488$ & $\mathbf{0.1035}$ & $         0.2253$ & $         0.2144$ & $         0.1441$ & $         0.0461$ & $        -0.0574$ & $-0.1453$ & $-0.1916$ & $-0.1517$ \\
$0.2364$ & $\mathbf{0.0705}$ & $         0.1783$ & $         0.2120$ & $         0.1619$ & $         0.0750$ & $        -0.0238$ & $-0.1120$ & $-0.1629$ & $-0.1301$ \\
$0.2239$ & $\mathbf{0.0490}$ & $\mathbf{0.1004}$ & $         0.1889$ & $         0.1636$ & $         0.0900$ & $        -0.0015$ & $-0.0871$ & $-0.1390$ & $-0.1095$ \\
$0.2115$ & $\mathbf{0.0336}$ & $\mathbf{0.0636}$ & $\mathbf{0.1334}$ & $         0.1544$ & $         0.0971$ & $         0.0146$ & $-0.0664$ & $-0.1174$ & $-0.0890$ \\
$0.1990$ & $\mathbf{0.0223}$ & $\mathbf{0.0404}$ & $\mathbf{0.0715}$ & $         0.1338$ & $         0.0980$ & $         0.0266$ & $-0.0484$ & $-0.0971$ & $-0.0680$ \\
$0.1866$ & $\mathbf{0.0140}$ & $\mathbf{0.0246}$ & $\mathbf{0.0406}$ & $\mathbf{0.0769}$ & $         0.0924$ & $         0.0350$ & $-0.0325$ & $-0.0777$ & $-0.0456$ \\
$0.1742$ & $\mathbf{0.0079}$ & $\mathbf{0.0136}$ & $\mathbf{0.0217}$ & $\mathbf{0.0364}$ & $         0.0777$ & $         0.0398$ & $-0.0185$ & $-0.0588$ & $-0.0194$ \\
$0.1617$ & $\mathbf{0.0036}$ & $\mathbf{0.0062}$ & $\mathbf{0.0097}$ & $\mathbf{0.0154}$ & $\mathbf{0.0272}$ & $         0.0399$ & $-0.0062$ & $-0.0399$ &           \\
$0.1493$ & $\mathbf{0.0010}$ & $\mathbf{0.0017}$ & $\mathbf{0.0027}$ & $\mathbf{0.0041}$ & $\mathbf{0.0066}$ & $\mathbf{0.0127}$ & $ 0.0043$ & $-0.0200$ &           \\
\end{tabular}
\end{ruledtabular}
\end{table*}
\endgroup

\begingroup
\begin{table}
\caption{\label{t:a1km3} Radiative corrections $(\alpha/\pi)A_n$ and $(\alpha/\pi)A_{nF}$ in the TBR and FBR for $K^0\to \pi^- + \mu^+ + \nu_\mu$ decay. The entries are (a) $(\alpha/\pi)A_1 \times 10$, (b) $(\alpha/\pi)A_2\times 10^2$, and (c) $(\alpha/\pi)A_3\times 10^3$, and those corresponding to the FBR are marked in boldface characters. The energies $E$ and $E_2$ are given in $\textrm{GeV}$.}
\begin{ruledtabular}
\begin{tabular}{lrrrrrrrrr}
$E_2$\textbackslash$E$ & $0.1131$ & $ 0.1280$ & $0.1429$ & $0.1578$ & $0.1727$ & $0.1876$ & $0.2025$ & $0.2174$ & $0.2323$ \\ 
\hline
(a)      &                    &                   &                   &                   &                   &           &           &           &           \\
$0.2480$ &                    &                   &                   & $        -0.0405$ & $        -0.0228$ & $-0.0200$ & $-0.0207$ & $-0.0207$ & $-0.0144$ \\
$0.2361$ &                    & $         0.0513$ & $         0.0490$ & $         0.0398$ & $         0.0272$ & $ 0.0134$ & $ 0.0001$ & $-0.0101$ & $-0.0110$ \\
$0.2242$ & $          0.0846$ & $         0.0808$ & $         0.0709$ & $         0.0573$ & $         0.0411$ & $ 0.0239$ & $ 0.0073$ & $-0.0058$ & $-0.0088$ \\
$0.2123$ & $          0.1005$ & $         0.0890$ & $         0.0775$ & $         0.0632$ & $         0.0464$ & $ 0.0283$ & $ 0.0107$ & $-0.0035$ & $-0.0069$ \\
$0.2004$ & $          0.1384$ & $         0.0910$ & $         0.0777$ & $         0.0637$ & $         0.0474$ & $ 0.0296$ & $ 0.0121$ & $-0.0021$ & $-0.0054$ \\
$0.1885$ & $\mathbf{ 0.0089}$ & $         0.1009$ & $         0.0748$ & $         0.0610$ & $         0.0458$ & $ 0.0290$ & $ 0.0123$ & $-0.0013$ &           \\
$0.1766$ & $\mathbf{ 0.0046}$ & $\mathbf{0.0111}$ & $         0.0742$ & $         0.0562$ & $         0.0422$ & $ 0.0269$ & $ 0.0115$ & $-0.0011$ &           \\
$0.1647$ & $\mathbf{ 0.0022}$ & $\mathbf{0.0050}$ & $\mathbf{0.0095}$ & $         0.0513$ & $         0.0370$ & $ 0.0236$ & $ 0.0099$ & $-0.0144$ &           \\
$0.1528$ & $\mathbf{ 0.0008}$ & $\mathbf{0.0018}$ & $\mathbf{0.0031}$ & $\mathbf{0.0060}$ & $         0.0308$ & $ 0.0190$ & $ 0.0073$ &           &           \\
$0.1409$ & $\mathbf{ 0.0000}$ & $\mathbf{0.0001}$ & $\mathbf{0.0003}$ & $\mathbf{0.0006}$ & $\mathbf{0.0012}$ & $ 0.0126$ & $-0.0006$ &           &           \\
(b)      &                    &                   &                       &                       &                   &           &           &           &           \\
$0.2480$ &                    &                   &                   & $        -0.0380$ & $        -0.0227$ & $-0.0198$ & $-0.0194$ & $-0.0186$ & $-0.0153$ \\
$0.2361$ &                    & $         0.0575$ & $         0.0428$ & $         0.0268$ & $         0.0124$ & $ 0.0003$ & $-0.0095$ & $-0.0164$ & $-0.0207$ \\
$0.2242$ & $          0.1494$ & $         0.1072$ & $         0.0746$ & $         0.0487$ & $         0.0275$ & $ 0.0099$ & $-0.0044$ & $-0.0157$ & $-0.0269$ \\
$0.2123$ & $          0.2290$ & $         0.1475$ & $         0.1007$ & $         0.0669$ & $         0.0402$ & $ 0.0183$ & $ 0.0003$ & $-0.0148$ & $-0.0358$ \\
$0.2004$ & $          0.4634$ & $         0.2020$ & $         0.1304$ & $         0.0863$ & $         0.0536$ & $ 0.0274$ & $ 0.0057$ & $-0.0135$ & $-0.0576$ \\
$0.1885$ & $\mathbf{ 0.0096}$ & $         0.3368$ & $         0.1730$ & $         0.1103$ & $         0.0693$ & $ 0.0380$ & $ 0.0121$ & $-0.0121$ &           \\
$0.1766$ & $\mathbf{ 0.0037}$ & $\mathbf{0.0144}$ & $         0.2635$ & $         0.1449$ & $         0.0893$ & $ 0.0508$ & $ 0.0196$ & $-0.0114$ &           \\
$0.1647$ & $\mathbf{ 0.0011}$ & $\mathbf{0.0052}$ & $\mathbf{0.0180}$ & $         0.2107$ & $         0.1177$ & $ 0.0671$ & $ 0.0285$ & $-0.0144$ &           \\
$0.1528$ & $\mathbf{-0.0001}$ & $\mathbf{0.0014}$ & $\mathbf{0.0055}$ & $\mathbf{0.0198}$ & $         0.1674$ & $ 0.0895$ & $ 0.0380$ &           &           \\
$0.1409$ & $\mathbf{-0.0002}$ & $\mathbf{0.0000}$ & $\mathbf{0.0008}$ & $\mathbf{0.0026}$ & $\mathbf{0.0079}$ & $ 0.1218$ & $-0.0077$ &           &           \\
(c)      &                    &                   &                       &                       &                   &           &           &           &           \\
$0.2480$ &                    &                   &                   & $        -0.0135$ & $        -0.0145$ & $-0.0177$ & $-0.0213$ & $-0.0254$ & $-0.0325$ \\
$0.2361$ &                    & $         0.0155$ & $         0.0059$ & $        -0.0034$ & $        -0.0119$ & $-0.0200$ & $-0.0284$ & $-0.0387$ & $-0.0615$ \\
$0.2242$ & $          0.0767$ & $         0.0509$ & $         0.0314$ & $         0.0154$ & $         0.0015$ & $-0.0117$ & $-0.0257$ & $-0.0441$ & $-0.0911$ \\
$0.2123$ & $          0.1610$ & $         0.1033$ & $         0.0699$ & $         0.0451$ & $         0.0242$ & $ 0.0047$ & $-0.0162$ & $-0.0448$ & $-0.1312$ \\
$0.2004$ & $          0.4178$ & $         0.1883$ & $         0.1258$ & $         0.0869$ & $         0.0564$ & $ 0.0287$ & $-0.0009$ & $-0.0424$ & $-0.2234$ \\
$0.1885$ & $\mathbf{ 0.0106}$ & $         0.3884$ & $         0.2114$ & $         0.1447$ & $         0.9936$ & $ 0.0606$ & $ 0.0200$ & $-0.0384$ &           \\
$0.1766$ & $\mathbf{ 0.0079}$ & $\mathbf{0.0247}$ & $         0.3832$ & $         0.2294$ & $         0.1570$ & $ 0.1019$ & $ 0.0466$ & $-0.0360$ &           \\
$0.1647$ & $\mathbf{ 0.0065}$ & $\mathbf{0.0165}$ & $\mathbf{0.0413}$ & $         0.3824$ & $         0.2386$ & $ 0.1557$ & $ 0.0788$ & $-0.0468$ &           \\
$0.1528$ & $\mathbf{ 0.0050}$ & $\mathbf{0.0114}$ & $\mathbf{0.0223}$ & $\mathbf{0.0552}$ & $         0.3751$ & $ 0.2284$ & $ 0.1138$ &           &           \\
$0.1409$ & $\mathbf{ 0.0017}$ & $\mathbf{0.0036}$ & $\mathbf{0.0064}$ & $\mathbf{0.0120}$ & $\mathbf{0.0270}$ & $ 0.3285$ & $-0.0231$ &           &           \\
\end{tabular}
\end{ruledtabular}
\end{table}
\endgroup

\section{\label{sec:vus}Inroads in determining $V_{us}$}

The present result (\ref{eq:dgfinal}) completes a calculational program for obtaining the RC to the DP of all $K_{l3}$ modes ($K = K^\pm, K^0$, $l=e,\mu$) to order $(\alpha/\pi)(q/M_1)$ in a model-independent fashion. A compact expression can thus be written as
\begin{equation}
d\Gamma(K_{l3}) = C_K^2 \frac{G_F^2}{32\pi^3}|V_{us}|^2M_1^3 dEdE_2 \left[ \mathcal{A}_0 + \frac{\alpha}{\pi} \left(\mathcal{A}_{\mathrm{TBR}} + \mathcal{A}_{\mathrm{FBR}} \right) \right], \label{eq:dfinal}
\end{equation}
where $C_K$ is a Clebsch-Gordan coefficient ($C_K$ is $1/\sqrt{2}$ and 1 for charged and neutral kaons, respectively) and $\mathcal{A}_{\mathrm{0}}$, $\mathcal{A}_\mathrm{TBR}$, and $\mathcal{A}_{\mathrm{FBR}}$ are functions of the kinematical variables and depend quadratically on the form factors; the subscripts X attached to these $\mathcal{A}$'s denote the corresponding contribution to the DP: the 0 subscript refers to the uncorrected contribution whereas TBR and FBR are self-descriptive subscripts to denote contributions arising from RC. The quantity $\mathcal{A}_0$ can be found in Ref.~\cite{jl2011}. The quantities $\mathcal{A}_\mathrm{TBR}$ and $\mathcal{A}_\mathrm{FBR}$ are provided in Refs.~\cite{jl2011} and \cite{jl2012} for charged kaons, respectively, and in Ref.~\cite{jl2015} and the present analysis for neutral kaons, respectively. The different $\mathcal{A}_{\mathrm X}$ can be generically written as 
\begin{equation}
\mathcal{A}_\mathrm{X} = A_1 |f_+(q^2)|^2 + A_2 \mathrm{Re} \,[f_+(q^2) f_-^*(q^2)] + A_3 |f_-(q^2)|^2, \label{eq:ax}
\end{equation}
where the $A_i$ amplitudes are straightforwardly obtained for the three cases as $\mathcal{A}_\mathrm{X}$ are.

The primary aim of constructing Eq.~(\ref{eq:dfinal}) is to achieve a reliable determination of $|V_{us}|$. For this purpose, an important matter is the unambiguous determination of all the inputs, in particular, the dependence on the square of the four-momentum transfer of the form factors. Quite often, analyses of $K_{\mu 3}$ data assume a linear dependence of $f_\pm$ on $q^2$, namely \cite{particle},
\begin{equation}
f_\pm(q^2) = f_\pm(0) \left [1 + \lambda_\pm \frac{q^2}{M_2^2} \right],
\end{equation}
where $\lambda_\pm$ are the average slopes of the form factors. For most $K_{\mu 3}$ data, a constant $f_-$ suffices.

Recent analyses have introduced the form factors $f_+$ and $f_0$, which are associated with vector and scalar exchange, respectively, to the lepton pair. In terms of $f_\pm$, $f_0$ is
\begin{equation}
f_0(q^2) = f_+(q^2) + \frac{q^2}{M_1^2-M_2^2}f_-(q^2),
\end{equation}
or equivalently,
\begin{equation}
f_-(q^2) = \frac{M_1^2-M_2^2}{q^2} \left[ f_0(q^2)-f_+(q^2) \right].
\end{equation}

For $f_-$ constant, $f_0$ becomes
\begin{equation}
f_0(q^2) =  f_0(0) \left [1 + \lambda_0 \frac{q^2}{M_2^2} \right].
\end{equation}

A common practice advocated recently by high-statistics experiments consists in including a quadratic term in the expansion of $f_+(q^2)$, namely \cite{particle},
\begin{equation}
f_+(q^2) = f_+(0) \left[1 + \lambda_+^\prime \frac{q^2}{M_2^2} + \frac12 \lambda_+^{\prime\prime} \left(\frac{q^2}{M_2^2} \right)^2 \right],
\end{equation}
where $\lambda_+^\prime$ and $\lambda_+^{\prime\prime}$ are the slope and curvature of the form factor, respectively.

The differential decay rate in terms of the form factors $f_+$ and $f_0$ is thus
\begin{equation}
d\Gamma(K_{l3}) = C_K^2 \frac{G_F^2}{32\pi^3}|V_{us}|^2M_1^3 dEdE_2 \left[ \mathcal{B}_0 + \frac{\alpha}{\pi} \left(\mathcal{B}_{\mathrm{TBR}} + \mathcal{B}_{\mathrm{FBR}} \right) \right],
\end{equation}
where
\begin{equation}
\mathcal{B}_\mathrm{X} = B_1 |f_+(q^2)|^2 + B_2 \mathrm{Re} \,[f_+(q^2) f_0^*(q^2)] + B_3 |f_0(q^2)|^2. \label{eq:bx}
\end{equation}
The $B_i$ amplitudes are related to the $A_i$ ones through
\begin{equation}
B_1 = A_1 - \frac{M_1^2-M_2^2}{q^2} A_2 + \left[ \frac{M_1^2-M_2^2}{q^2} \right]^2 A_3,
\end{equation}
\begin{equation}
B_2 = \frac{M_1^2-M_2^2}{q^2} A_2 - 2 \left[ \frac{M_1^2-M_2^2}{q^2} \right]^2 A_3,
\end{equation}
\begin{equation}
B_3 = \left[ \frac{M_1^2-M_2^2}{q^2} \right]^2 A_3.
\end{equation}

Now, integrating $d\Gamma(K_{l3})$ over the energies of the emitted charged lepton and pion, restricted to the appropriate kinematical region, yields the total decay rate $\Gamma(K_{l3})$; it can be written as
\begin{equation}
\Gamma(K_{l3}) = C_K^2 \frac{G_F^2}{128\pi^3}M_1^5 \left[ f_+^{K^0\pi^-}(0) |V_{us}|\right]^2 S_{\mathrm{ew}} I(\lambda_+^\prime,\lambda_+^{\prime\prime},\lambda_0), \label{eq:dtotal}
\end{equation}
where $S_{\mathrm{ew}}=1.0232(3)$ is the short-distance electroweak correction. Notice that the form factor $f_+^{K^0\pi^-}(0)$ is used to normalize the form factors of all channels. Isospin-breaking corrections are thus accounted for in the factor $\Delta_\mathrm{SU(2)}^{K\pi}$, which is defined by
\begin{equation}
\Delta_\mathrm{SU(2)}^{K^0\pi^-} = 0, \qquad \qquad \Delta_\mathrm{SU(2)}^{K^+\pi^0} = \frac{f_+^{K^+\pi^0}(0)}{f_+^{K^0\pi^-}(0)} -1.
\end{equation}

A detailed analysis \cite{anto} yields
\begin{equation}
\Delta_\mathrm{SU(2)}^{K^+\pi^0} = 0.027 \pm 0.004,
\end{equation}
which is the value actually used here.

On the other hand, the function $I(\lambda_+^\prime,\lambda_+^{\prime\prime},\lambda_0)$ introduced in Eq.~(\ref{eq:dtotal}) can be worked out to get
\begin{equation}
I(\lambda_+^\prime,\lambda_+^{\prime\prime},\lambda_0) =
h_0
+ h_1 \lambda_+^\prime 
+ h_2 {\lambda_+^\prime}^2
+ h_3 \lambda_+^{\prime\prime} 
+ h_4 \lambda_+^\prime  \lambda_+^{\prime\prime} 
+ h_5 {\lambda_+^{\prime\prime}}^2
+ h_6 \lambda_0
+ h_7 \lambda_+^\prime \lambda_0 
+ h_8 \lambda_+^{\prime\prime} \lambda_0
+ h_9 \lambda_0^2. \label{eq:ii}
\end{equation}

For numerical purposes, the values of slopes and curvatures of the form factors to be used here are the ones determined in Ref.~\cite{anto}, namely,
\begin{eqnarray}
&  & \lambda_+^\prime = (25.02 \pm 1.12)\times 10^{-3}, \qquad \lambda_+^{\prime\prime} = (1.57 \pm 0.48)\times 10^{-3}, \qquad \lambda_0 = (13.34 \pm 1.41)\times 10^{-3}, \label{eq:lms}
\end{eqnarray}
and $\mu$-$e$ universality will be assumed throughout the analysis.

The various $h_m$ coefficients in Eq.~(\ref{eq:ii}) are made up of three terms, which arise from the different contributions to the DP, in the same spirit as the $\mathcal{A}_\mathrm{X}$ quantities discussed above; explicitly,
\begin{equation}
h_m = h_m^{(0)} + \frac{\alpha}{\pi} \left( h_m^\mathrm{TBR} + h_m^\mathrm{FBR} \right). \label{eq:iij}
\end{equation}

The generic quantities $h_m^\mathrm{X}$ are defined by
\begin{equation}
h_0 = \frac{4}{M_1^2} \int_{\mathcal{D}} dEdE_2 \, A_1,
\end{equation}
\begin{equation} 
h_1 = \frac{4}{M_1^2} \int_{\mathcal{D}} dEdE_2 \left[ 2A_1 \frac{q^2}{M_2^2} + A_2 \left[1-\frac{M_1^2}{M_2^2}\right] \right],
\end{equation}
\begin{equation}
h_2 = \frac{4}{M_1^2} \int_{\mathcal{D}} dEdE_2 \left[ A_1 \left[ \frac{q^2}{M_2^2} \right]^2 + A_2 \frac{q^2}{M_2^2} \left[1-\frac{M_1^2}{M_2^2}\right] + A_3 \left[1-\frac{M_1^2}{M_2^2}\right]^2 \right],
\end{equation}
\begin{equation}
h_3 = \frac{4}{M_1^2} \int_{\mathcal{D}} dEdE_2 \, \frac{q^2}{M_2^2} \left[ A_1 \frac{q^2}{M_2^2} + \frac12 A_2 \left[1-\frac{M_1^2}{M_2^2}\right] \right],
\end{equation}
\begin{equation}
h_4 = \frac{4}{M_1^2} \int_{\mathcal{D}} dEdE_2 \, \frac{q^2}{M_2^2} \left[ A_1 \left[ \frac{q^2}{M_2^2} \right]^2 + A_2 \frac{q^2}{M_2^2} \left[ 1-\frac{M_1^2}{M_2^2} \right] + A_3 \left[ 1-\frac{M_1^2}{M_2^2} \right]^2 \right],
\end{equation}
\begin{equation}
h_5 = \frac{4}{M_1^2} \int_{\mathcal{D}} dEdE_2 \left[ \frac{q^2}{M_2^2} \right]^2 \left[ \frac14 A_1 \left[ \frac{q^2}{M_2^2} \right]^2 + \frac14 A_2 \frac{q^2}{M_2^2} \left[1-\frac{M_1^2}{M_2^2}\right] + \frac14 A_3 \left[1-\frac{M_1^2}{M_2^2} \right]^2 \right],
\end{equation}
\begin{equation}
h_6 = \frac{4}{M_1^2} \int_{\mathcal{D}} dEdE_2 \, A_2 \left[-1+\frac{M_1^2}{M_2^2}\right],
\end{equation}
\begin{equation}
h_7 = \frac{4}{M_1^2} \int_{\mathcal{D}} dEdE_2 \left[1-\frac{M_1^2}{M_2^2}\right] \left[ -A_2 \frac{q^2}{M_2^2} - 2 A_3 \left[ 1-\frac{M_1^2}{M_2^2} \right] \right],
\end{equation}
\begin{equation}
h_8 = \frac{4}{M_1^2} \int_{\mathcal{D}} dEdE_2 \, \frac{q^2}{M_2^2} \left[1-\frac{M_1^2}{M_2^2}\right] \left[ -\frac12 A_2 \frac{q^2}{M_2^2} - A_3 \left[1-\frac{M_1^2}{M_2^2} \right] \right],
\end{equation}
\begin{equation}
h_9 = \frac{4}{M_1^2} \int_{\mathcal{D}} dEdE_2 \, A_3 \left[ 1-\frac{M_1^2}{M_2^2} \right]^2.
\end{equation}

The region of integration $\mathcal{D}$ in the TBR is delimited by
\begin{equation}
E_2^{\mathrm{min}} \leq E_2 \leq E_2^{\textrm{max}}, \qquad \quad m \leq E \leq E_m, \label{eq:EE2lim}
\end{equation}
where
\begin{equation}
E_2^{\mathrm{max,min}} = \frac12 (M_1-E \pm l) + \frac{M_2^2}{2(M_1-E\pm l)},
\end{equation}
and
\begin{equation}
E_m = \frac{1}{2M_1} (M_1^2-M_2^2+m^2), \label{eq:Elim}
\end{equation}
whereas $\mathcal{D}$ in the FBR is delimited by
\begin{equation}
M_2 \leq E_2 \leq E_2^{\textrm{min}}, \qquad \quad m \leq E \leq E_c,
\end{equation}
where
\begin{equation}
E_c = \frac12 (M_1-M_2) + \frac{m^2}{2(M_1-M_2)}.
\end{equation}

The $h_m$ coefficients can be computed once and for all by means of numerical integrations; they are listed in Tables \ref{t:hsc} and \ref{t:hsn} for the $K_{l3}^\pm$ and $K_{l3}^0$ processes, respectively, for the three contributions of interest. Let us notice that there are large cancellations between the TBR and FBR contributions in the charged sector, being more noticeable in the electron mode.

\begingroup
\begin{table}
\caption{\label{t:hsc}Numerical values of the coefficients $h_m^{(0)}$, $(\alpha/\pi)h_m^\mathrm{TBR}$, and $(\alpha/\pi)h_m^\mathrm{FBR}$ introduced in Eq.~(\ref{eq:iij}) for the process $K_{l3}^\pm$.
}
\begin{tabular}{lrrrrrr}
\hline\hline
& \multicolumn{3}{c}{$K_{e3}^\pm$} & \multicolumn{3}{c}{$K_{\mu 3}^\pm$} \\[1mm]
& (0) & TBR & FBR & (0) & TBR & FBR \\
\hline
$h_0$ & $0.096532$ & $-0.000739$ & $0.000469$ & $0.062307$ & $-0.000130$ & $ 0.000006$ \\
$h_1$ & $0.356832$ & $-0.003679$ & $0.003112$ & $0.215480$ & $-0.000187$ & $ 0.000033$ \\
$h_2$ & $0.527921$ & $-0.006989$ & $0.006225$ & $0.372524$ & $-0.000647$ & $ 0.000097$ \\
$h_3$ & $0.527921$ & $-0.006989$ & $0.006225$ & $0.372524$ & $-0.000322$ & $ 0.000089$ \\
$h_4$ & $1.949234$ & $-0.030664$ & $0.027891$ & $1.473792$ & $-0.002450$ & $ 0.000536$ \\
$h_5$ & $2.051334$ & $-0.036613$ & $0.033706$ & $1.608508$ & $-0.002684$ & $ 0.000756$ \\
$h_6$ &            &             &            & $0.151085$ & $-0.000601$ & $ 0.000035$ \\
$h_7$ &            &             &            & $0.000000$ & $ 0.000649$ & $-0.000015$ \\
$h_8$ &            &             &            & $0.000000$ & $ 0.001043$ & $-0.000045$ \\
$h_9$ &            &             &            & $0.284003$ & $-0.001594$ & $ 0.000112$ \\
\hline\hline
\end{tabular}
\end{table}

\begingroup
\begin{table}
\caption{\label{t:hsn}Numerical values of the coefficients $h_m^{(0)}$, $(\alpha/\pi)h_m^\mathrm{TBR}$, and $(\alpha/\pi)h_m^\mathrm{FBR}$ introduced in Eq.~(\ref{eq:iij}) for the process $K_{l3}^0$.
}
\begin{tabular}{lrrrrrr}
\hline\hline
& \multicolumn{3}{c}{$K_{e3}^0$} & \multicolumn{3}{c}{$K_{\mu 3}^0$} \\[1mm]
& (0) & TBR & FBR & (0) & TBR & FBR \\
\hline
$h_0$ & $0.093890$ & $-0.000006$ & $0.000512$ & $0.060571$ & $0.000549$ & $ 0.000016$ \\
$h_1$ & $0.324506$ & $ 0.000667$ & $0.003246$ & $0.195891$ & $0.002710$ & $ 0.000126$ \\
$h_2$ & $0.448458$ & $ 0.000473$ & $0.006157$ & $0.316290$ & $0.004073$ & $ 0.000325$ \\
$h_3$ & $0.448458$ & $ 0.000473$ & $0.006157$ & $0.316290$ & $0.005091$ & $ 0.000313$ \\
$h_4$ & $1.545880$ & $-0.000586$ & $0.026069$ & $1.168170$ & $0.017015$ & $ 0.001688$ \\
$h_5$ & $1.518290$ & $-0.002770$ & $0.029702$ & $1.189850$ & $0.018751$ & $ 0.002240$ \\
$h_6$ &            &             &            & $0.137012$ & $0.001354$ & $ 0.000030$ \\
$h_7$ &            &             &            & $0.000000$ & $0.002044$ & $-0.000024$ \\
$h_8$ &            &             &            & $0.000000$ & $0.003606$ & $-0.000088$ \\
$h_9$ &            &             &            & $0.240366$ & $0.001834$ & $ 0.000088$ \\
\hline\hline
\end{tabular}
\end{table}

Using the $h_m$ coefficients, the functions $I(\lambda_+^\prime,\lambda_+^{\prime\prime},\lambda_0)$ can be obtained for the different decay modes using the values provided in Eq.~(\ref{eq:lms}). These functions are made up of three terms, namely, $I^{(0)} + I^\mathrm{TBR} + I^\mathrm{FBR}$, and they are listed in Table \ref{t:is}.

\begingroup
\begin{table}
\caption{\label{t:is}Values of the function $I(\lambda_+^\prime,\lambda_+^{\prime\prime},\lambda_0)=I^{(0)} + I^\mathrm{TBR} + I^\mathrm{FBR}$ as introduced in Eq.~(\ref{eq:dtotal}).
}
\begin{tabular}{lrrr}
\hline\hline
Process & $I^{(0)}$ & $I^\mathrm{TBR}\times 10^4$ & $I^\mathrm{FBR}\times 10^4$ \\
\hline
$K_{e3}^+$    &  $0.106701$ & $-8.477080$ & $5.612508$ \\
$K_{\mu 3}^+$ &  $0.070644$ & $-1.440362$ & $0.077138$ \\
$K_{e3}^0$    &  $0.103059$ & $ 0.112902$ & $6.079239$ \\
$K_{\mu 3}^0$ &  $0.068086$ & $ 6.472170$ & $0.203570$ \\
\hline\hline
\end{tabular}
\end{table}

At this stage, all the required inputs to Eq.~(\ref{eq:dtotal}) are now defined, so the only unknown left, $f_+^{K^0\pi^-}(0)|V_{us}|$, can be obtained from experimental data \cite{particle} through a least-squares fit. The measured quantities are the kaon lifetimes and branching ratios, straightforwardly converted into the decay rates of $K_{l3}^\pm$ and $K_{l3}^0$, the latter given for $K_L$ and $K_S$. This information is listed in Table \ref{t:data} for the sake of completeness.

\begingroup
\begin{table}
\caption{\label{t:data}Decay rates obtained from experimental data on $K_{l3}$ decays reported in Ref.~\cite{particle}, $\Gamma_\mathrm{exp}$, and their theoretical estimates, $\Gamma_\mathrm{th}$. Uncorrected, TBR, FBR, and isospin-breaking contributions that make up $\Gamma_\mathrm{th}$ are also displayed. The contributions to $\chi^2$ of each mode coming from the fit are listed in the last column. The units of $\Gamma$ are $10^6 \, \mathrm{s}^{-1}$.
}
\begin{tabular}{lcrrrrrr}
\hline\hline
Process & $\Gamma_\mathrm{exp}$ & $\Gamma_\mathrm{th}$ & $\Gamma^{(0)}$ & $\Delta \Gamma^\mathrm{TBR}$ & $\Delta \Gamma^\mathrm{FBR}$ & $\Delta \Gamma^\mathrm{SU(2)}$ & $\Delta \chi^2$ \\
\hline
$K^+ \to \pi^0 e^+ \nu_e$           & $4.095 \pm 0.033$ & $4.119$ & $3.918$ & $-0.031$ & $0.020$ & $0.212$ & $0.55$ \\
$K^+ \to \pi^0 \mu^+ \nu_\mu$       & $2.708 \pm 0.028$ & $2.729$ & $2.594$ & $-0.005$ & $0.000$ & $0.140$ & $0.58$ \\
$K_L^0 \to \pi^\pm e^\mp \nu_e$     & $7.926 \pm 0.039$ & $7.923$ & $7.876$ & $ 0.001$ & $0.046$ &         & $0.01$ \\
$K_L^0 \to \pi^\pm \mu^\mp \nu_\mu$ & $5.285 \pm 0.026$ & $5.254$ & $5.203$ & $ 0.049$ & $0.002$ &         & $1.41$ \\
$K_S^0 \to \pi^\pm e^\mp \nu_e$     & $7.862 \pm 0.089$ & $7.923$ & $7.876$ & $ 0.001$ & $0.046$ &         & $0.47$ \\
$K_S^0 \to \pi^\pm \mu^\mp \nu_\mu$ & $5.238 \pm 0.056$ & $5.254$ & $5.203$ & $ 0.049$ & $0.002$ &         & $0.08$ \\
\hline\hline
\end{tabular}
\end{table}

Without further ado, a global fit to data yields
\begin{equation}
f_+^{K^0\pi^-}(0) |V_{us}| = 0.2168 \pm 0.0003, \label{eq:myproduct}
\end{equation}
with $\chi^2=3.1$ for five degrees of freedom. It is important to remark that the quoted error comes from the fit only and does not include any theoretical uncertainty. The best-fit value Eq.~(\ref{eq:myproduct}) is in very good agreement to the values $0.21673 \pm 0.00059$ and $0.2163 \pm 0.0005$ presented in Refs.~\cite{ambro} and \cite{anto}, respectively. The former was obtained primarily from $K_L$ mesons while the latter was obtained within a global analysis of leptonic and semileptonic data.

Recent lattice QCD estimates of $f_+^{K^0\pi^-}(0)$ can be found in the literature. A selection of them can be found in Refs.~\cite{kane,primer,boy}. Using these estimates, the corresponding values of $|V_{us}|$ are provided in Table \ref{t:vus}. All in all, these values are in complete agreement with the one suggested in Ref.~\cite{particle}.

\begingroup
\begin{table}
\caption{\label{t:vus}Values of form factor $f_+^{K^0\pi^-}(0)$ as determined by recent lattice analyses and the corresponding value of $|V_{us}|$ obtained via Eq.~(\ref{eq:myproduct}).
}
\begin{tabular}{lcc}
\hline\hline
Collaboration & $f_+^{K^0\pi^-}(0)$ & $V_{us}$ \\
\hline
JLQCD \cite{kane}                       &  $0.959 \pm 0.008$  & $0.2261 \pm 0.0019$ \\
Fermilab Lattice and MILC \cite{primer} & $0.9704 \pm 0.0033$ & $0.2234 \pm 0.0018$ \\
RBC/UKQCD \cite{boy}                    & $0.9685 \pm 0.0037$ & $0.2239 \pm 0.0009$ \\
\hline\hline
\end{tabular}
\end{table}

Some interesting features of the analysis can be better appreciated in Table \ref{t:data}. For each decay mode, apart from $\Gamma_\mathrm{exp}$, the predicted decay rate $\Gamma_\mathrm{th}$ along with the contributions that constitute it are also listed. RC represent a small fraction of the total decay rate; they amount to $-0.26\%$, $-0.18\%$, $0.60\%$, and $0.98\%$ for $K_{e3}^+$, $K_{\mu 3}^+$, $K_{e3}^0$, and $K_{\mu 3}^0$ decays, respectively. Isospin-breaking corrections, on the other hand, play a leading role in the first two processes because they represent a non-negligible $5.14\%$ of $\Gamma_\mathrm{th}$.

\section{\label{sec:summ}Closing remarks}

As it was pointed out in the introductory section, the present paper completes a calculational program of RC to the DP of $K_{l3}$ decays to order $(\alpha/\pi)(q/M_1)$, for both charged and neutral kaons and whether the emitted charged lepton is an electron or a muon. The effects of the so-called three- and four-body regions of this DP have been accounted for. 

It has been shown that the analysis of RC in $K_{l3}$ decays can be made in a general unambiguous fashion by extending the gauge-invariant approach originally introduced by Sirlin \cite{sirlin} to study the RC to neutron decay. On general grounds, the RC are separated into model-independent and model-dependent parts. The former is finite in the ultraviolet and fully contains the infrared divergence. The latter can be totally absorbed into the already existing strong-interaction form factors, without introducing new ones. The price for this is that the measured form factors will be not the pure strong-interaction form factors but new ones which are modified by the model dependence of RC. This is not a drawback, however, because strictly speaking it is only these modified form factors that can actually be experimentally determined. The resolution of which part of them is due to strong interactions only and which part belongs to RC is a theoretical issue.

Following the lines of previous works \cite{jl2011,jl2012,jl2015}, in the present analysis the FBR contribution of the RC to the DP of neutral kaons is obtained. The first result is provided by identifying all the triple integrals over the bremsstrahlung photon, which could be considered as final if numerical integrations are performed. However, it is also possible to proceed further and perform such integrals analytically, which constitutes the second result of this work. This leaves the complete calculation on the same footing as the ones of previous works \cite{jl2011,jl2012,jl2015}.

Indeed, the most important contribution of this paper is to show how to incorporate the full RC of order $(\alpha/\pi)(q/M_1)$ to the DP of $K_{l3}$ towards a reliable determination of $|V_{us}|$. The analysis is performed by comparing the theoretical expression for the decay rate against the experimental information available \cite{particle}. The comparison is made through a least-squares fit for the only parameter left, namely, $f_+^{K^0\pi^-}(0)|V_{us}|$. The fit yields $\chi^2=3.1$ for five degrees of freedom for $f_+^{K^0\pi^-}(0) |V_{us}| = 0.2168 \pm 0.0003$, which can be compared to the values determined in Refs.~\cite{ambro,anto}.

Recent lattice QCD estimates of $f_+^{K^0\pi^-}(0)$ can be found in the literature \cite{kane,primer,boy}, which would allow one to evaluate $|V_{us}|$. With the advent of more refined simulations, or even new or improved data the accuracy in $|V_{us}|$ can be also improved. The RC presented here and in previous works \cite{jl2011,jl2012,jl2015} can be useful in such a task.

\begin{acknowledgments}
The authors are grateful to Consejo Nacional de Ciencia y Tecnolog{\'\i}a (Mexico) for partial support. A.M., C.J.-L., and J.J.T.\ were partially supported by Comisi\'on de Operaci\'on y Fomento de Actividades Acad\'emicas (Instituto
Polit\'ecnico Nacional). R.F.-M.\ was also partially supported by Fondo de Apoyo a la Investigaci\'on (Universidad Aut\'onoma de San Luis Potos{\'\i}).
\end{acknowledgments}

\end{document}